\documentclass[preprint,aps,nofootinbib]{revtex4}

\usepackage{graphicx}
\usepackage{epsfig,amssymb,amsmath,color}
\usepackage{epstopdf}

\def\m{\mu}

\newcommand{\beq}{\begin{equation}}
\newcommand{\eeq}{\end{equation}}
\newcommand{\bea}{\begin{eqnarray}}
\newcommand{\eea}{\end{eqnarray}}
\newcommand{\bear}{\begin{array}}
\newcommand {\eear}{\end{array}}
\newcommand{\bef}{\begin{figure}}
\newcommand {\eef}{\end{figure}}
\newcommand{\bec}{\begin{center}}
\newcommand {\eec}{\end{center}}

\newcommand{\la}{\left\langle}
\newcommand{\ra}{\right\rangle}
\newcommand{\ds}{\displaystyle}

\def\EQ#1{Eq.~(\ref{#1})}

\def\REF#1{(\ref{#1})}
\def\GEV#1{10^{#1}{\rm\,GeV}}

\def\lrfp#1#2#3{ \left(\frac{#1}{#2} \right)^{#3}}
\def\oten#1{ {\mathcal O}(10^{#1})}

\begin{document}
\draft
\tighten
\preprint{UT-14-16, TU-965, IPMU14-0092
}
\title{\large \bf
Gravitino Problem in Supergravity Chaotic Inflation \\
and SUSY Breaking Scale  after BICEP2
}
\author{
Kazunori Nakayama\,$^{a,\,c\,\ast}$\footnote[0]{$^\ast$ email: kazunori@hep-th.phys.s.u-tokyo.ac.jp},
Fuminobu Takahashi\,$^{b,\,c\,\dagger} $\footnote[0]{$^\star$ email: fumi@tuhep.phys.tohoku.ac.jp},
Tsutomu T. Yanagida\,$^{c\,\dagger} $\footnote[0]{$^\dagger$ email: tsutomu.yanagida@ipmu.jp}
    }
\affiliation{
    $^a$ Department of Physics, University of Tokyo, Tokyo 113-0033, Japan\\
    $^b$ Department of Physics, Tohoku University, Sendai 980-8578, Japan \\
    $^c$ Kavli IPMU, TODIAS, University of Tokyo, Kashiwa 277-8583, Japan
    }
%\date{\today}

\vspace{2cm}

\begin{abstract}
Gravitinos are generically produced by inflaton decays, which place  tight constraints on inflation models as well as 
supersymmetry breaking scale.  We revisit the gravitino production from decays of the inflaton and the 
supersymmetry breaking field, based on a chaotic inflation model suggested by the recent BICEP2 result. 
We study cosmological constraints on thermally and non-thermally produced gravitinos for a wide range of the gravitino mass,
and show that there are only three allowed regions of the gravitino mass: $m_{3/2}\lesssim 16$\,eV,
$m_{3/2}\simeq 10$--$1000$\,TeV and $m_{3/2} \gtrsim 10^{13}$\,GeV. 
\end{abstract}
\maketitle

%%%%%%%%%%%%%%%%%%%%%%%%%%%%%%%%%%%%%%%%%%%%
\section{Introduction}
%%%%%%%%%%%%%%%%%%%%%%%%%%%%%%%%%%%%%%%%%%%%

Recently the BICEP2 collaboration reported a detection of the primordial  B-mode polarization of 
the cosmic microwave background (CMB)~\cite{Ade:2014xna}, which, if confirmed, would provide
the strong case for inflation~\cite{Guth:1980zm,Linde:1981mu}.
The BICEP2 result can be explained by a large tensor-to-scalar ratio, $r=0.20^{+0.07}_{-0.05}$.
Taken at face value, it implies large-field inflation models, where the inflaton field excursion exceeds 
the Planck scale~\cite{Lyth:1996im}.

Among various large-field inflation models, by far the simplest one is the quadratic chaotic inflation model~\cite{Linde:1983gd}
given by
\begin{equation}
	V(\varphi) = \frac{1}{2}m^2\varphi^2,   \label{quad}
\end{equation}
where $\varphi$ is the inflaton, and  the inflaton mass is fixed to be  $m \simeq 2\times 10^{13}$\,GeV 
by the  the observed curvature perturbations. The energy density of the Universe during inflation is 
close to the GUT scale, and therefore, it is conceivable that the inflation model (\ref{quad}) is realized in the framework of supergravity
or string theory. The chaotic inflation model in supergravity was proposed in Ref.~\cite{Kawasaki:2000yn,Kawasaki:2000ws},
where an approximate shift symmetry on the inflaton was introduced  to have good control over inflaton field values 
greater than the Planck scale.\footnote{
There are various large-field inflation models in the supergravity  and superstring 
theory~\cite{Freese:1990ni,Murayama:1992ua,Dimopoulos:2005ac,Kallosh:2007ig,Silverstein:2008sg,McAllister:2008hb,Kaloper:2008fb,Takahashi:2010ky,
Nakayama:2010kt,Kallosh:2010ug,Nakayama:2010sk,Kallosh:2010xz,Harigaya:2012pg,Croon:2013ana,Nakayama:2013jka,
Nakayama:2013nya,Cicoli:2014sva,Czerny:2014wza,Czerny:2014xja,Nakayama:2014koa,Harigaya:2014sua,Harigaya:2014qza}. }

In order to lead to the standard big bang cosmology after inflation, 
the inflaton must transfer its energy to the standard model (SM) particles, i.e., reheating of the Universe. 
In supergravity, the inflaton  generically decays  through various Planck-suppressed interactions,
unless the inflaton is charged under unbroken symmetry. 
Specifically,  the inflaton decays into top quarks and Higgs, gluon pairs, right-handed neutrinos, etc., 
even without introducing ad hoc couplings with the visible sector~\cite{Endo:2006qk,Endo:2007ih,Endo:2006nj}. 
Some unwanted relics, however, are also
produced at the same time. One of such unwanted relics is the gravitino. In fact, it is known that 
gravitinos are generically produced by decays of the inflaton~\cite{Kawasaki:2006gs,Asaka:2006bv,Endo:2007ih,
Endo:2007sz} and the moduli~\cite{Endo:2006zj,Dine:2006ii,Endo:2006tf}, and the solutions to the gravitino
overproduction problem were studied in Refs.~\cite{Endo:2006xg,Endo:2007cu,Nakayama:2012hy}.

The amount of gravitinos produced by the inflaton decay depends on the properties of the inflaton and
the supersymmetry (SUSY) breaking field. The gravitino production rate  is enhanced
for a heavier inflaton with a larger coefficient of the linear term in the K\"ahler potential. In many inflation models, the latter
approximately coincides with the vacuum expectation value (VEV) of the inflaton or waterfall fields after inflation. 
The gravitino overproduction problem becomes acute especially if the SUSY breaking field, $z$,  is a purely singlet, 
i.e., the so called Polonyi field, as in the gravity mediation~\cite{Dine:2006ii,Endo:2006tf,Kawasaki:2006gs}. 
In this case the inflaton decay produces too many gravitinos, and as a result, 
various inflation models are tightly constrained or excluded for a wide range of the gravitino 
mass~\cite{Kawasaki:2006gs,Asaka:2006bv,Endo:2007ih,Endo:2007sz}.

Moreover, the Polonyi field itself causes a serious cosmological problem~\cite{Coughlan:1983ci};
the coherent oscillations of the Polonyi field easily dominate the energy density of the Universe,
and typically decay into the SM particles during and after the big-bang nucleosynthesis (BBN),
 altering the light element abundances in contradiction
with observations, and they also produce too many lightest SUSY particles (LSPs).
It is possible to consider dynamical SUSY breaking scenarios in which $z$ is charged under some symmetry and 
is stabilized with a heavy SUSY breaking mass~\cite{Banks:1993en,Coughlan:1983ci,Nakayama:2012hy,Evans:2013nka}. 
Then, the Polonyi problem becomes significantly 
relaxed since the Polonyi field can be stabilized at the enhanced symmetry point during  inflation, 
suppressing the initial oscillation amplitude.  Also, the inflaton decay rate into
 gravitinos and the Polonyi fields can be suppressed~\cite{Nakayama:2012hy}.

In this letter we revisit the gravitino overproduction problem in the chaotic inflation in light of the recent BICEP2 data,
for a wide range of the gravitino mass. In particular we take account of various sources for the gravitino production;
thermal production as well as  non-thermal one  from decays of the inflaton and the Polonyi field. We will show that there are
only three allowed regions of the gravitino mass, $m_{3/2}\lesssim 16$\,eV, $m_{3/2}\simeq 10$--$1000$\,TeV and 
$m_{3/2} \gtrsim 10^{13}$\,GeV.

%%%%%%%%%%%%%%%%%%%%%%%%%%%%%%%%%%%%%%%%%%%%
\section{Chaotic Inflation in Supergravity}
%%%%%%%%%%%%%%%%%%%%%%%%%%%%%%%%%%%%%%%%%%%%
In this section we briefly review the chaotic inflation model given in Ref.~\cite{Kawasaki:2000yn,Kawasaki:2000ws}.
We impose a shift symmetry on the inflaton superfield,
\begin{equation}
\label{shift}
	\phi \to \phi + iC,
\end{equation}
where $C$ is a real transformation parameter. With an abuse of notation, we
shall use the same symbol to denote both a chiral superfield and its scalar component, unless noted otherwise.
%Assuming that the shift symmetry is broken by the superpotential, 
The relevant interactions are given by
\begin{align}
	K_{\rm inf} &= c(\phi + \phi^\dagger) + \frac{1}{2}(\phi+\phi^\dagger)^2 + |X|^2 - k |X|^4 + \cdots  \label{K} \\
	W_{\rm inf} &= m X\phi,
\end{align}
where $c$ and $k$ are real constants of order unity and $X$ is a gauge singlet chiral superfield,
and we assume that $X$ has an $R$ charge $+2$ whereas $\phi$ is neutral.
Here we have chosen the origin of $\phi$ so that the superpotential  takes the above form.
The inflaton is identified with $\varphi \equiv \sqrt{2}{\rm Im}\,\phi$. 
The K\"ahler potential respects the shift symmetry (\ref{shift}), which is explicitly broken by  the superpotential.
Here and in what follows, we adopt the Planck unit in which the reduced Planck mass is set to be unity, $M_P=1$.

The scalar potential in supergravity is given by
\bea
V_{\rm sugra} = e^K\left((D_i W) K^{i {\bar j}} (D_j W)^* - 3 |W|^2\right).
\eea
The inflaton potential is given by  (\ref{quad}) even for $\varphi\gg 1$ because of the shift symmetry.
Note that, during inflation,  $X$ is stabilized at the origin $X=0$ for $k \gtrsim \mathcal O(1)$,
whereas the real component of $\phi$ is stabilized at ${\rm Re} \,\phi \approx -c/2$. After inflation, both $X$ and $\phi$
are stabilized at the origin.  For the graceful exit of the inflation we assume $c$ is at most of order unity~\cite{Kawasaki:2000ws}.
Note that the real component of $\phi$ starts to oscillate after inflation with an initial
amplitude $\sim c/2$. For $c = {\cal O}(1)$,  its abundance is comparable to that of the inflaton $\varphi$.

One can impose a discrete $Z_2$ symmetry under which both $\phi$ and $X$ flip their sign. Then those terms proportional to odd powers 
of $(\phi+\phi^\dag)$ in the K\"ahler potential are absent, and in particular, $c=0$. In this case, the inflaton decay into gravitinos can be forbidden.
On the other hand, it becomes non-trivial to induce the inflaton decay.  As long as the $Z_2$ symmetry
is unbroken, the inflaton would be stable unless we assign the $Z_2$ charge on the SM particles and its SUSY partners. 
Otherwise, we would need to break the $Z_2$ symmetry to induce the inflaton decay. Therefore, introducing a $Z_2$
symmetry imposes extra conditions on the structure of the underlying theory. 
Throughout this letter we assume that there is no such $Z_2$ symmetry
so that there is a linear term in the K\"ahler potential with $c = {\cal O}(1)$.

The inflaton decay into the visible sector in the chaotic inflation model was studied in detail in Ref.~\cite{Endo:2006nj}.
In the presence of the linear term in the K\"ahler potential, the inflaton is automatically coupled to all the fields that appear
in the superpotential with gravitational strength. For instance, the inflaton can decay into top (s)quarks and Higgs(ino) 
at tree level~\cite{Endo:2006qk}, and a pair of gauge bosons and gauginos at one-loop level~\cite{Endo:2007ih}.
Also the inflaton decays into a pair of the right-handed (s)neutrinos if kinematically allowed. The 
inflaton decay rate into the visible sector is of order $m^3$, and the reheating temperature 
is  $\oten{9}$\,GeV for $c = {\cal O}(1)$, without introducing ad hoc couplings with the visible sector.
This is one of the virtues of the chaotic inflation without the $Z_2$ symmetry. Therefore we will adopt the inflaton decay rate 
$\Gamma_{\rm inf} \sim 1$\,GeV  and the reheating temperature $T_{\rm R} \sim \GEV{9}$ as reference values
in the following analysis. 

Lastly let us comment on the mass eigenstates of the inflaton sector after inflation. We will add a constant term $W_0 \simeq m_{3/2}$ 
in the superpotential  in order to cancel positive contributions from the SUSY breaking. Then, $\phi$ and $X$ get maximally mixed with each other to form mass eigenstates~\cite{Kawasaki:2006gs}, 
 \bea
 \Phi_\pm \equiv \frac{ \phi \pm X^\dag}{\sqrt{2}}.
 \eea
Both eigenstates have a mass $\simeq m$. 
This mixing is effective only if the inflaton decay rate is smaller than the gravitino mass.
For our reference value of the inflaton decay rate, the critical value is  $m_{3/2} \sim 1$\,GeV.
For $m_{3/2} \gtrsim {\cal O}(1)$\,GeV,  the inflaton can decay through the interactions of $\phi$ and $X$ with
the other sectors, and as we shall see in the next section, the mixing between $\phi$ and $X$ is crucial for non-thermal 
gravitino production.  On the other hand, for $m_{3/2} \lesssim {\cal O}(1)$\,GeV, 
$\phi$ is an effective mass eigenstate until the decay,  the mixing between $\phi$ and $X$ is irrelevant
for the inflaton decay. In this case,  although the direct gravitino production from the inflaton 
becomes ineffective, too many gravitinos are thermally produced for $T_{\rm R} \sim \GEV{9}$ except
  $m_{3/2} \lesssim 16$\,eV~\cite{Viel:2005qj}.
In the next section we will study the gravitino production from various sources in detail.

%%%%%%%%%%%%%%%%%%%%%%%%%%%%%%%%%%%%%%%%%%%%
\section{Gravitino production}
%%%%%%%%%%%%%%%%%%%%%%%%%%%%%%%%%%%%%%%%%%%%
Gravitinos are produced from various sources. First, gravitinos are generically produced non-thermally
by inflaton decays. As we shall see shortly, the gravitino production rate sensitively depends on the
properties of the SUSY breaking field. Secondly, gravitinos are produced by thermal scatterings in plasmas.
Thirdly,  the Polonyi field mainly decays into a pair of gravitinos, if kinematically allowed. 
The Polonyi field is copiously produced by coherent oscillations, and it is also produced by the inflaton decays.
We will consider these production processes alternately. 

%%%%%%%%%%%%%%%%%%%%%%%%%%%%%%%%%%%%%%%%%%%%
\subsection{Non-thermal production from inflaton decays}
%%%%%%%%%%%%%%%%%%%%%%%%%%%%%%%%%%%%%%%%%%%%

%The Polonyi model for the SUSY breaking suffers from the serious cosmological problem~\cite{Coughlan:1983ci},
%and hence we consider the dynamical SUSY breaking model in which the SUSY breaking field obtains a large mass.
%Below the dynamical scale $\Lambda$, the inflaton and the SUSY breaking sector can be described as follows:

The gravitino production from inflaton decays proceeds through couplings between the inflaton and SUSY breaking
field(s). In the following we assume $m > 2m_{3/2}$. Otherwise gravitinos are not produced by the inflaton decay.

Let us add a simple extension of the Polonyi model to the inflation model \REF{K}:
\begin{equation}
\begin{split}
\label{lowE}
	K &= K_{\rm inf} + |z|^2 - \frac{|z|^4}{\Lambda^2}, \\
	W &= W_{\rm inf} + \mu^2 z + W_0,
\end{split}
\end{equation}
where $z$ is the SUSY breaking field, $\Lambda$ is an effective cut-off scale for the quartic
coupling, and $W_0$ is a constant term that is 
required to make the  cosmological constant (almost) zero in the present Universe.
The linear term of $z$ in the superpotential could be generated dynamically, and the quartic 
coupling in the K\"ahler potential may or may not be induced by the same dynamics. We 
however do not specify the origin of $\mu^2$ to retain generality, and indeed, the following discussion does not
depend on it.\footnote{Actually the situation gets worse if $\mu^2$ is of dynamical origin
and if its dynamical scale is below the inflaton mass, as the inflaton directly decays into hidden gauge sectors,
producing many gravitinos~\cite{Endo:2007ih}. }  For simplicity we take $\mu^2$ and $W_0$ real in the
following.

Let us first consider the original Polonyi model which is obtained in the limit of $\Lambda \to \infty$. 
In this case $z$ is a purely singlet, and there is no special point in the field space. 
One can show that $z$ is stabilized at $\la z \ra = \sqrt{3}-1$ with the $F$-term, $F^z = - \sqrt{3} \m^2$.
In the original Polonyi model, there is the notorious cosmological Polonyi problem~\cite{Coughlan:1983ci}. 
The Polonyi field is copiously produced by 
coherent oscillations, and easily dominate the Universe after reheating. For the gravitino mass comparable to 
or lighter than ${\cal O}(1)$\,TeV, it decays during or after BBN, altering the light element abundances in 
contradiction with observations. In fact, even if the Polonyi problem is solved by some mechanism\footnote{
One possibility is the so called adiabatic suppression mechanism~\cite{Linde:1996cx,Takahashi:2010uw,Nakayama:2011wqa}.
}, there is the gravitino overproduction from the inflaton decay. As the SUSY breaking field $z$ is a purely singlet,
the following operator is allowed~\cite{Dine:2006ii,Endo:2006tf,Kawasaki:2006gs},
\bea
\label{phizz}
K_{\rm int} &=& \frac{1}{2} (\phi+ \phi^\dag) zz + {\rm h.c.}.
\eea
This leads to the coupling of the inflaton to goldstinos $\chi$,
\bea
{\cal L} &=& m X \chi \chi + {\rm h.c.} = m \frac{\Phi_+^\dag - \Phi_-^\dag}{\sqrt{2}} \chi \chi + {\rm h.c.} 
\eea
where we have expanded the $F$-term of the inflaton in terms of $X$ and used $m \approx W_{\phi X}$,
and $\chi$ is the fermionic component of $z$. Therefore, the inflaton mass eigenstates $\Phi_{\pm}$ decays
into a pair of gravitinos with a rate of order $m^3$, which results in the gravitino overproduction for a wide
range of the gravitino mass. Here and in what follows,  we assume that 
 the mixing between $\phi$ and $X$ is effective as long as the non-thermal gravitino
production from inflaton decay is concerned.
% except for $m_{3/2} \lesssim 16$\,eV and $m_{3/2} \gtrsim \GEV{13}$.

The Polonyi problem can be ameliorated if the SUSY breaking field $z$ is charged under some symmetry
as in the dynamical SUSY breaking scenario~\cite{Banks:1993en}. The low-energy effective theory is given
by \EQ{lowE} with the cut-off scale $\Lambda \ll 1$. The $F$-term of $z$ is given by $F_z \approx
 - \mu^2 = \sqrt{3}m_{3/2}$, and  the mass  is given by
\begin{equation}
	m_z^2 \simeq \frac{12m_{3/2}^2}{\Lambda^2}.
\end{equation}
The low energy true minimum  is located at 
$$\langle z\rangle \simeq 2\sqrt{3} \lrfp{m_{3/2}}{m_z}{2} \simeq \frac{m_{3/2}}{m_z} \Lambda.$$
During inflation the $z$ can be stabilized in the vicinity of
 the origin where the symmetry is enhanced, suppressing the initial oscillation amplitude. Still, some amount of
 coherent oscillations of $z$ is induced because the low-energy minimum of $z$ is slightly deviated from the origin.
We will return to the (relaxed) Polonyi problem later in this section.

Let us estimate the gravitino production from inflation decays in the case of $\Lambda \ll 1$. 
Since $z$ is charged under some symmetry, interactions like \REF{phizz} and 
$K_{\rm int} = (\phi+\phi^\dagger) z + {\rm h.c.}$ are forbidden. Note however that the effective interactions
are induced from higher order terms, since $z$ develops a small but non-zero VEV. For instance let us consider
\bea
K &=& (\phi+\phi^\dag) \frac{|z|^4}{\Lambda^2},
\eea
which leads to the effective interaction (\ref{phizz}) with a coefficient suppressed by $\la z \ra^2/\Lambda^2 \sim m_{3/2}^2/m_z^2$. 
The contribution
to the gravitino pair production rate is therefore
%%%
%\bea
%\Delta \Gamma(\Phi \to \psi_{3/2} \psi_{3/2}) & \simeq&
%% \frac{c'^2}{256 \pi} \la z \ra^4 \frac{m^3}{M_P^2}
%\frac{ c'^2}{8 \pi} \lrfp{m_{3/2}}{m_z}{4}  \frac{m^3}{M_P^2},
%\eea
%%%
%which is 
significantly suppressed for $m_z \gg m_{3/2}$, as expected. 

There is another process that becomes relevant for a heavy $m_z$, which can be understood as follows. 
First,  the SUSY breaking field $z$ is known to mainly decay into a pair of gravitinos,
as its coupling is enhanced for longitudinal modes (see \EQ{zgg}). This is an analogue of the standard model Higgs boson
which would mainly decay into the longitudinal modes of $W$ bosons for the Higgs mass heavier than twice the
$W$ boson mass. Secondly, there is a non-zero mixing  between $X$ and $z$ in supergravity.
Thus, the inflaton decays into a pair of gravitinos through the mixing between $X$ and $z$.

To get the feeling of how the decay proceeds, let us write down a part of the interactions leading to the mixing;
\bea
V_{\rm sugra} \supset e^K K_\phi W K^{\phi {\bar \phi}} (D_\phi W) + {\rm h.c.} \supset m \la K_\phi \ra \mu^2 z X^\dag  + {\rm h.c.},
\eea
where we have expanded $W$ and $W_\phi$ with respect to $z$ and $X$, respectively.
This induces a mixing between $z$ and $X$, and the mixing angle $\theta$ depends on the relation between $m$ and $m_z$;
\bea
\theta \sim \left\{
\bear{cc}
\ds{\frac{m_{3/2}}{m} K_\phi} & {\rm~for~}m \gg m_z,\\
&\\
\ds{\frac{m_{3/2} m }{m_z^2} K_\phi }& {\rm~for~}m \ll m_z.\\
\eear
\right.
\eea
There is another contribution to the mixing of comparable size from interaction like $K = (\phi+\phi^\dag) |z|^2$. 
Although not shown here, there is also a contribution from the kinetic mixing, $K_{\phi {\bar z}} \sim \la z \ra$.
The following results on the gravitino production rate
can be understood by combining the above mixing angle and the decay rate of $z$ into a pair of gravitinos given in \EQ{zgg}.
The detailed expressions for the gravitino pair production rate are given in Ref.~\cite{Endo:2006tf}.

Here we summarize the gravitino production rate (cf. Ref.~\cite{Nakayama:2012hy});
\begin{equation}
	\Gamma(\Phi \to 2\psi_{3/2}) \simeq 
	\begin{cases}
%		\displaystyle
%		\frac{d^2}{64\pi}\left( \frac{m_z}{m} \right)^4\left( \frac{\langle K_\phi\rangle}{M_P} \right)^2 \frac{m^3}{M_P^2}
%		& {\rm~for~}m \gg m_z,\\
%		\displaystyle
%		\frac{d^2}{64\pi}\left( \frac{\langle K_\phi\rangle}{M_P} \right)^2 \frac{m^3}{M_P^2}
%		& {\rm~for~}m \ll m_z,
		\displaystyle
		\frac{ 1}{8 \pi}\frac{m^3}{M_P^2} \left(
		 {\tilde c}^2 \lrfp{m_{3/2}}{m}{2} +  c'^2\left( \frac{m_{3/2}}{m_z} \right)^4+\frac{d^2}{8} \left( \frac{m_z}{m} \right)^4
		 \right)
		& {\rm~for~}m \gg m_z,\\
%		  \frac{m^3}{M_P^2}
%		\frac{ {\tilde c}^2}{8 \pi} \lrfp{m_{3/2}}{m}{2}  \frac{m^3}{M_P^2}+
%		\frac{ c'^2}{8 \pi} \lrfp{m_{3/2}}{m_z}{4}  \frac{m^3}{M_P^2}+
%		\frac{d^2}{64\pi}\left( \frac{m_z}{m} \right)^4 \frac{m^3}{M_P^2}
%		& {\rm~for~}m \gg m_z \gg m_{3/2},\\
		\displaystyle
		\frac{d^2}{64\pi}  \frac{m^3}{M_P^2}
		& {\rm~for~}m \ll m_z,
	\end{cases}
	\label{PhiSS}
\end{equation}
where we have defined\footnote{
Precisely speaking, ${\tilde c}$ should be defined in the  eigenstate  basis 
of the non-analytic mass terms~\cite{Endo:2006zj}. Our results are not changed by this simplification. 
}
\bea
%	d\langle K_\phi\rangle \equiv \left<  K_\phi - K_{\phi z\bar z} \right>.
	{\tilde c} &\equiv&  \la K_{\phi z {\bar z}} \ra,\\
	c' &\equiv& \frac{1}{4} \la K_{\phi z {\bar z}z {\bar z}} \ra,\\
	d &\equiv& \left<  K_\phi - K_{\phi z\bar z} \right>,
\eea
which are considered to be of order unity.
The second term in $d$ has a comparable contribution to the first term if there is an interaction 
like $K \supset (\phi+\phi^\dagger)|z|^2$.

The inflaton also decays into a pair of the SUSY breaking fields with the rate
\begin{equation}
	\Gamma(\Phi \to zz^\dagger) = 
%	\frac{{\tilde d}^2}{32\pi}\left( \frac{m_z}{m} \right)^4
%	\left( \frac{\langle K_\phi\rangle}{M_P} \right)^2 \frac{m^3}{M_P^2}
%	\left( 1- \frac{4m_z^2}{m^2} \right)^{1/2}  {\rm~for~}m > 2m_z,
	\frac{{\tilde d}^2}{32\pi}\left( \frac{m_z}{m} \right)^4
	 \frac{m^3}{M_P^2}
	\left( 1- \frac{4m_z^2}{m^2} \right)^{1/2}  {\rm~for~}m > 2m_z,
	\label{phizszs}
\end{equation}
where we have defined
\begin{equation}
	{\tilde d} \equiv \left<  K_\phi - K_{\phi z\bar z} + \frac{K_{z\bar zz\bar z\phi}}{K_{z\bar zz\bar z}} \right>.
\end{equation}
The second and third terms in ${\tilde d}$ have comparable contributions to the first term if there are operators like 
$K \supset (\phi+\phi^\dagger)|z|^2$ and   $(\phi+\phi^\dagger)|z|^4/\Lambda^2$. Since $z$ predominantly decays into a pair of gravitinos, this decay process
gives a  comparable contribution to the final gravitino abundance, with respect to the direct production.
We will set $c' = {\tilde c} = d={\tilde d}=1$ in the next section.

%
%If the inflaton is heavier than the dynamical scale $\Lambda$, it decays into hidden sector hadrons
%and hidden hadrons finally decay and produce gravitinos. The rate is given by~\cite{Endo:2007ih,Endo:2007sz}.
%%
%\begin{equation}
%	\Gamma(\Phi \to {\rm hadron}) =
%	\begin{cases}
%		\displaystyle
%		\frac{N_g\alpha_h^2}{512\pi^3}(\mathcal T_G-\mathcal T_R)^2\left( \frac{\langle K_\phi\rangle}{M_P} \right)^2 \frac{m^3}{M_P^2}
%		& {\rm~for~}m \gtrsim 2\Lambda,\\
%		\displaystyle
%		0
%		& {\rm~for~}m \lesssim 2\Lambda,
%	\end{cases}
%\end{equation}
%%
%where $\mathcal T_G$ and $\mathcal T_R$ are the Dynkin index of the adjoint representation and and matter fields in the representation $R$,
%respectively, $\alpha_h$ is the fine structure constant of the hidden gauge group and $N_g$ is the number of generators of the gauge group.

From these decay rates, it is obvious that the gravitino production rate is significantly suppressed for a certain range of
$m_z$ satisfying  $m_{3/2} \ll m_z \ll m$,
because of the suppression factor of $ (m_z/m)^4$ and $(m_{3/2}/m_z)^4$.
The non-thermal gravitino abundance is then given by
\begin{equation}
	Y_{3/2}^{(\phi)} =  \frac{3T_{\rm R}}{4m}\frac{2\Gamma(\Phi \to \tilde z\tilde z) +
	4\Gamma(\Phi \to zz^\dagger) }{\Gamma_{\rm tot}},
\end{equation}
where $\Gamma_{\rm tot} =(\pi^2 g_*/90)^{1/2}T_{\rm R}^2/M_P$ is the total decay rate of the inflaton.

%%%%%%%%%%%%%%%%%%%%%%%%%%%%%%%%%%%%%%%%%%%%
\subsection{Thermal production}
%%%%%%%%%%%%%%%%%%%%%%%%%%%%%%%%%%%%%%%%%%%%
Gravitinos are also produced by scatterings of SM particles and their superpartners in thermal bath.
The gravitino abundance is proportional to the reheating temperature $T_{\rm R}$~\cite{Bolz:2000fu}:
\begin{equation}
	Y_{3/2}^{\rm (th)} \simeq 
	\begin{cases}
		\displaystyle
		{\rm min}\left[2\times 10^{-12}\left(1+ \frac{m_{\tilde g}^2}{3m_{3/2}^2} \right)
		\left( \frac{T_{\rm R}}{10^{10}\,{\rm GeV}} \right), ~\frac{0.42}{g_{*s}(T_{3/2})} \right]  
		&{\rm~for~} T_{\rm R} \gtrsim m_{\rm SUSY}, \\
		\displaystyle
		0  &{\rm~for~} T_{\rm R} \lesssim m_{\rm SUSY}, \\
	\end{cases}
\end{equation}
where $m_{\tilde g}$ denotes the gluino mass, $m_{\rm SUSY}$ is the typical soft SUSY breaking mass, 
and $g_{*s}(T_{3/2})$ is the effective degrees of freedom at the gravtino freezeout, if the gravitinos are thermalized~\cite{Pierpaoli:1997im,Viel:2005qj}.
We will take $m_{\tilde g} = m_{\rm SUSY}$ in the next section.

%%%%%%%%%%%%%%%%%%%%%%%%%%%%%%%%%%%%%%%%%%%%
\subsection{Production from decays of the Polonyi field}
%%%%%%%%%%%%%%%%%%%%%%%%%%%%%%%%%%%%%%%%%%%%

The true minimum of the Polonyi field $z$ is deviated from the position during inflation,
and hence it starts to oscillate when the Hubble parameter becomes equal to the Polonyi mass.

If $m_z \ll m$, the Polonyi can be stabilized in the vicinity of the origin $z\sim 0$ due to the Hubble-induced mass term.
The low energy true minimum, on the other hand, is located at $\langle z\rangle = 2\sqrt{3}m_{3/2}^2/m_z^2$.
Thus it starts to oscillate with an amplitude of $\langle z\rangle$ when the Hubble parameter becomes comparable to $m_z$.
On the other hand,  the coherent oscillations are not induced if $m_z \gg m$,\footnote{$m$ is close to the Hubble scale at the end of inflation in the chaotic inflation model.} since the Polonyi adiabatically follows the temporal minimum of the potential 
in this case~\cite{Nakayama:2011wqa}.
Therefore, the abundance of Polonyi coherent oscillations is estimated as
\begin{equation}
	\frac{\rho_z}{s} \simeq
	\begin{cases}
		\displaystyle
		3T_{\rm R}\left(\frac{m_{3/2}}{m_z} \right)^4
		& {\rm~for~} m_z \ll m,\\
		\displaystyle
		0
		& {\rm~for~}m_z \gg m.
	\end{cases}
\end{equation}
The Polonyi field decays into  gravitinos with a rate:
\begin{equation}
\label{zgg}
	\Gamma(z\to 2 \psi_{3/2}) \simeq \frac{1}{96\pi} \frac{m_z^5}{m_{3/2}^2M_P^2}.
\end{equation}
%%
%Hence it may also cause the gravitino overproduction.
The gravitino abundance from the Polonyi decay is estimated as
\begin{equation}
	Y_{3/2}^{(z)} \simeq \frac{2}{m_z}\frac{\rho_z}{s}.
\end{equation}
We will take into account those contributions to the gravitino abundance in the next section.

%%%%%%%%%%%%%%%%%%%%%%%%%%%%%%%%%%%%%%%%%%%%
\section{SUSY breaking scale inferred from BICEP2}
%%%%%%%%%%%%%%%%%%%%%%%%%%%%%%%%%%%%%%%%%%%%

The gravitino abundance is given by the sum of all the contributions considered in the previous section;
\begin{equation}
	Y_{3/2} = Y_{3/2}^{\rm (th)}+Y_{3/2}^{(\phi)}+Y_{3/2}^{(z)}.
\end{equation}
Cosmological effects of the gravitino depend on its mass.
If it is stable, the energy density of the gravitino should not exceed the observed dark matter abundance.
If it is unstable and if its lifetime is longer than $\sim 1$\,sec, the gravitino abundance is severely constrained by BBN.
If its lifetime is much shorter than 1 sec, the LSPs produced by the gravitino decay is constrained by the dark matter abundance.
For concreteness, we assume the anomaly mediation for the  mass spectrum of the gauginos for $m_{3/2} > 100$\,TeV
and hence the Wino is the LSP.\footnote{
	We assume the conservation of R-parity.
	If the R-parity is  violated, the LSP can decay well before BBN and constrains can be significantly relaxed
	for $m_{3/2} \gg 10$\,TeV.
}
Also we assume the gauge mediation model for $m_{3/2} \lesssim 100$\,GeV in which the NLSP mass is around 1\,TeV,
so that the decay of NLSP is severely constrained by BBN if $m_{3/2} \gtrsim 1$\,GeV.

\begin{figure}[t]
\begin{center}
\includegraphics[scale=1.4]{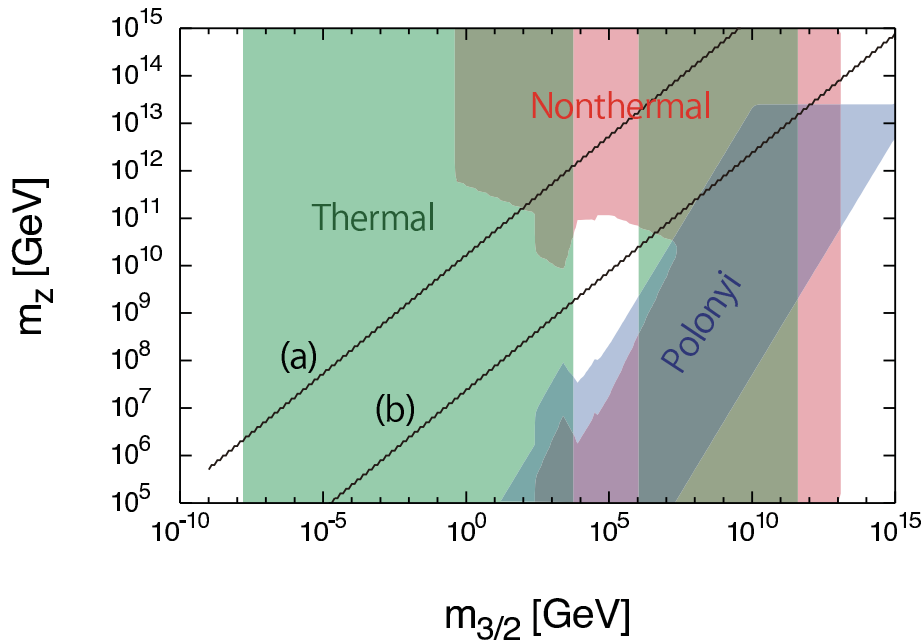}
\caption { 
The green shaded regions, purple shaded region and blue shaded region are excluded from 
thermal production ($Y_{3/2}^{\rm (th)}$), non-thermal production by the inflaton decay ($Y_{3/2}^{(\phi)}$)
and the Polonyi decay ($Y_{3/2}^{(z)}$), respectively.
The two black lines show typical values of $m_z$ (see text).
In this figure we have taken $T_{\rm R}=10^{9}$\,GeV.
}
\label{fig:TR1e9}
\end{center}
\end{figure}

Figure~\ref{fig:TR1e9} shows the constraints from the gravitino overproduction on the $(m_{3/2}, m_z)$ plane.
In this figure we have taken $T_{\rm R}=10^{9}$\,GeV,  for which thermal leptogenesis is possible~\cite{Fukugita:1986hr}.
	Note that the spontaneous inflaton decay through the top Yukawa coupling leads to
	$T_{\rm R}\sim 10^8$\,GeV. This is considered to be the lower bound on the reheating temperature~\cite{Endo:2006qk}.
The green, purple and blue shaded regions are excluded by thermal production ($Y_{3/2}^{\rm (th)}$),
non-thermal production by the inflaton decay ($Y_{3/2}^{(\phi)}$), and  the Polonyi decay ($Y_{3/2}^{(z)}$),
respectively. 
We consider non-thermal production of gravitinos from inflaton decays
only for $m_{3/2} \geq 1$\,GeV, since otherwise the mixing between $\phi$
and $X$ becomes ineffective.\footnote{
Still, there are other production processes which are not suppressed for $m_{3/2} \lesssim 1$\,GeV.
For instance, the abundance of the real component of $\phi$ is sizable and it decays into the lowest components of $z$
with a rate comparable to (\ref{phizszs}). This however does not change our conclusion, because thermally produced gravitinos
exceed the observed dark matter abundance in this case for $T_{\rm R} \sim \GEV{9}$. 
}
The two black lines show typical values of $m_z$.
If the mass of $z$ is generated radiatively, $\Lambda$ in (\ref{K}) is written as 
$\Lambda = (4\pi/\lambda^2) \Lambda_{\rm dyn}$ with $\Lambda_{\rm dyn}$ being the dynamical SUSY breaking scale, 
and $\lambda$ the coupling constant between $z$ and the strong dynamical sector.\footnote{
$\lambda$ corresponds to the coupling between $z$ and dynamical quarks 
in the IYIT model~\cite{Izawa:1996pk} (see Ref.~\cite{Nakayama:2012hy}).
}
The upper line (a) corresponds to the strongly coupled case, $\lambda=4\pi$ and the lower line (b) corresponds to
the weak coupling case $\lambda=1$. 
% (a) $m_z \simeq 9\sqrt{m_{3/2}M_P}$ and  (b) $0.02\sqrt{m_{3/2}M_P}$.

One can see clearly from the figure that there are only three allowed regions: 
%\begin{itemize}
%\item[(i)] $m_{3/2} \lesssim 16$\,eV 
%\item[(ii)] $m_{3/2}\sim 10-1000$\,TeV
%\item[(iii)] $m_{3/2} \gtrsim 10^{13}$\,GeV
%\end{itemize}
(i) $m_{3/2} \lesssim 16$\,eV, (ii) $m_{3/2}\sim 10-1000$\,TeV and (iii) $m_{3/2} \gtrsim 10^{13}$\,GeV.
Interestingly, the allowed region (ii)  is consistent with the pure gravity mediation 
scenario~\cite{Ibe:2011aa}/ the minimal split SUSY~\cite{ArkaniHamed:2012gw,Arvanitaki:2012ps},  
which naturally explains the 125\,GeV Higgs boson observed at LHC~\cite{Aad:2012tfa}.
If the SUSY breaking sector is truly strongly coupled, i.e. $\lambda = 4\pi$ corresponding to the line labeled by (a),
there are only two regions (i) and (iii). The second region (ii) is viable if the coupling of the SUSY breaking field $z$ is
perturbative, e.g. $\lambda = 1$ (the line labeled by (b)).

%%%%%%%%%%%%%%%%
%%%%%%%%%%%%%%%%

%%%%%%%%%%%%%%%%%%%%%%%%%%%%%%%%%%%%%%%%%%%%
\section{Conclusions}
%%%%%%%%%%%%%%%%%%%%%%%%%%%%%%%%%%%%%%%%%%%%

In this paper we revisited the gravitino overproduction problem in the chaotic inflationary Universe scenario,
in light of recent BICEP2 result.
Taking account of the non-thermal gravitino production from the direct inflaton decay as well as thermal production,
and also the effect of Polonyi coherent oscillation,
we have shown that there are only three allowed regions of the gravitino mass: $m_{3/2}\lesssim 16$\,eV,
$m_{3/2}\simeq 10$--$1000$\,TeV and $m_{3/2} \gtrsim 10^{13}$\,GeV.
It is interesting that, except for the trivial limits of ultra light and ultra heavy gravitino,
the gravitino mass of $\sim 100$\,TeV appeared from these considerations, which fits the pure gravity mediation scenario.
Interestingly, the inflaton decays into the visible sector even without introducing ad hoc couplings, because there is generically
a linear term of the inflaton in the K\"ahler potential. Therefore the inflaton generically decays into all the fields that appear in the superpotential, and  the reheating temperature is naturally as high as $\sim 10^{9}$\,GeV so that thermal leptogenesis successfully works.
The non-thermal leptogenesis is also possible, if the right-handed neutrino mass is close to the inflaton mass~\cite{Endo:2006nj}. 

The large tensor-to-scalar ratio observed by BICEP2 indicates a detectable level of stochastic gravitational wave background
of the primordial origin around the frequency of $\sim 1$\,Hz, which can be detected by future space-based gravitational wave detectors.
In particular, the observation of the shape of the gravitational wave spectrum enables us to determine the reheating temperature,
if it is around $\sim 10^9$\,GeV~\cite{Seto:2003kc,Nakayama:2008wy}.

%%%%%%%%%%%%%%%%%%%%%%%%%%%%%%%%%%%%%%%%%%%%
\section*{Acknowledgments}
%%%%%%%%%%%%%%%%%%%%%%%%%%%%%%%%%%%%%%%%%%%%
This work was supported by the Grant-in-Aid for Scientific Research on
Innovative Areas (No.23104008 [FT]),  JSPS Grant-in-Aid for
Young Scientists (B) (No.24740135) [FT], and Inoue Foundation for Science [FT].  This work was also
supported by World Premier International Center Initiative (WPI Program), MEXT, Japan.

%%%%%%%%%%%%%%%%%%%%%%%%%%%%%%%%%%%%%%%%%%%%%%%%

\end{document}